\def\be{\begin{equation}}
\def\ee{\end{equation}}
\def\ba{\begin{array}}
\def\ea{\end{array}}

\documentclass[aps,amsmath,amssymb,amsfonts,showpacs,column,pra]{revtex4}
\usepackage{graphicx}
\usepackage{epstopdf}
\def\qed{\leavevmode\unskip\penalty9999 \hbox{}\nobreak\hfill
     \quad\hbox{\leavevmode  \hbox to.77778em{%
               \hfil\vrule   \vbox to.675em%
               {\hrule width.6em\vfil\hrule}\vrule\hfil}}
     \par\vskip3pt}

\begin{document}
\title{ Analytical Expression of Quantum Discord for Rank-2 Two-qubit States}
\author{Xue-Na Zhu$^{1}$}
\author{Shao-Ming Fei$^{2,3}$}
\thanks{feishm@cnu.edu.cn}
\author{Xianqing Li-Jost$^{3,4}$}

\affiliation{$^1$School of Mathematics and Statistics Science, Ludong University, Yantai 264025, China\\
$^2$School of Mathematical Sciences, Capital Normal University
Beijing 100048, China\\
$^3$Max-Planck-Institute for Mathematics in the Sciences, 04103
Leipzig, Germany\\
$^4$School of Mathematics and Statistics, Hainan
Normal University, Haikou, 571158, China
}

\begin{abstract}

Quantum correlations characterized by quantum entanglement and quantum discord play important roles in
many quantum information processing.
We study the relations among the entanglement of formation, concurrence, tangle, linear entropy based  classical correlation and
von Neumann entropy based  classical correlation . We present analytical formulae of linear entropy based classical correlation for arbitrary $d\otimes 2$ quantum states and  von Neumann entropy based classical correlation for arbitrary $2\otimes 2$ rank-2 quantum states.
From the von Neumann entropy based classical correlation, we derive an explicit formula of quantum discord for arbitrary rank-2 two-qubit quantum states.
\end{abstract}

\maketitle

\section{Introduction}
Correlations between the subsystems of a bipartite system
play significant roles in many information processing tasks and physical
processes.
The quantum entanglement \cite{R.Horodecki} is an important kind of quantum correlation which
plays significant roles in many quantum
tasks such as quantum teleportation, dense coding, swapping, error correction and
remote state preparation.
A bipartite state is called separable if it has zero
entanglement between subsystems $A$ and $B$:
the probabilities of the measurement outcomes from measuring the subsystem $A$ are independent of the
probabilities of the measurement outcomes from measuring the subsystem $B$.
Nevertheless, a separable state may still have quantum correlation -- quantum discord, if it is
impossible to learn all the mutual information by measuring one of the subsystems.
Quantum discord is the minimum amount of correlation, as measured by mutual information,
that is necessarily lost in a local measurement of bipartite quantum states.
It has been shown that the quantum discord is required for some information processing like
assisted optimal state discrimination \cite{roa,libo1}.

Let $\rho_{AB}$ denote the density operator of a bipartite system $H_A\otimes H_B$.
The quantum mutual information is defined by
\begin{equation}\label{total}
 I(\rho_{AB})=S(\rho_A)+S(\rho_B)-S(\rho_{AB}),
\end{equation}
where $\rho_{A(B)}=Tr_{B(A)}(\rho_{AB})$ are reduced density matrices,
$S(\rho)=-Tr(\rho\log\rho)$ is the Von Neumann
entropy. Quantum mutual information is the information-theoretic measure of the total
correlation in bipartite quantum states. In terms of measurement-based conditional density operators,
the classical correlation of bipartite states $\rho_{AB}$ is defined by \cite{prl88017901},
\begin{equation}\label{classical}
 I^{\leftarrow}(\rho_{AB})=\max_{\{P_i\}}[S(\rho_A)-\sum_ip_iS(\rho^i_A)],
\end{equation}
where the maximum is taken over all positive operator-valued measure (POVM)
${P_i}$ performed on subsystem $B$, satisfying $\sum_iP_i^{\dagger}P_i=I$ with probability
of $i$ as an outcome, $p_i=Tr[(I_A\otimes P_i)\rho_{AB}(I_A\otimes P_i^{\dagger})],$
$\rho^i_A=Tr_B[(I_A\otimes P_i)\rho_{AB}(I_A\otimes P_i^{\dagger})]/p_i$ is the conditional
states of system A associated with outcome $i$, $I_A$ and $I$ are the corresponding identity operators.

The quantum discord is defined as the difference between the total
correlation (\ref{total}) and the classical correlation (\ref{classical}) \cite{prl88017901,jpa}:
\begin{equation}\label{discord}
 Q^{\leftarrow}(\rho_{AB})=I(\rho_{AB})-I^{\leftarrow}(\rho_{AB}).
\end{equation}

Generally it is a challenging problem to compute the quantum correlation
$Q^{\leftarrow}(\rho_{AB})$ due to difficulty in computing the
classical correlation $I^{\leftarrow}(\rho_{AB})$. Analytically
formulae of $\mathcal{Q}(\rho)$ can be obtained only for some special quantum states like
Bell-diagonal states \cite{luoshunlong}, X-type states \cite{libo} with respect to projective measurements, as well as some special two-qubit states \cite{pra77042303}.
In stead of analytical formulae, some estimation on the lower and upper bounds of quantum discord
are also obtained \cite{pra84052112,zhihaoma}. A lower bound of quantum discord for the 2-qutrit systems
is obtained in \cite{c1}.
In \cite{MP} a hierarchy of computationally efficient lower bounds to the standard quantum discord
has been presented.

In this paper, by studying the classical correlations of $d\otimes2$  quantum states,
we present the analytical formula of quantum discord for any two-qubit states with rank-2.

To derive an analytical formula of quantum discord for rank-2 two-qubit states
under von Neumann entropy, we first study the classical correlation under
linear entropy. The linear entropy $S_2(\rho)$ of a quantum state $\rho$ is given by
$S_2(\rho)=2[1-Tr(\rho^2)]$.
The linear entropy version of the classical correlation (\ref{classical}) of a bipartite state $\rho_{AB}$ is given by $I^{\leftarrow}_2(\rho_{AB})=\max[S_2(\rho_A)-\sum_ip_iS_2(\rho^i_A)]$.

Any $d\otimes2$ bipartite  quantum state $\rho_{AB}$ may be written as\cite{prl96220503}
\begin{equation}\label{VB}
 \rho_{AB}=\varLambda_{\rho}\otimes I_B(|r_{B^{\prime}B}\rangle\langle r_{B^{\prime}B}|),
\end{equation}
where $|r_{B^{\prime}B}\rangle$ is the symmetric two qubit purification of the
reduced density operator $\rho_{B}$ on an auxiliary qubit system $B^{\prime}$ and $\varLambda_{\rho}$
is a qubit channel from  $B^{\prime}$ to $A$.

A qudit states can be written as the Bloch expression $\rho=\frac{I_d+\vec{r}\gamma}{d}$, where
$I_d$ denotes the $d\times d$ identity matrix, $\vec{r}$ is a $d^2-1$
dimensional real vector, $\gamma=(\lambda_1,\lambda_2,...,\lambda_{d^2-1})^{T}$ is the vector
of the generators of $SU(d)$ and $T$ stands for transpose.
The linear entropy written in terms of the Bloch vector $\vec{r}$ of a qudit state, is given by
$S_2(\frac{I_d+\vec{r}\gamma}{d})=\frac{2d^2-2d-4|\vec{r}|^2}{d^2}$.
The action of a qubit channel $\varLambda$ on a single-qubit state
$\rho=\frac{I_2+\vec{r}_B\sigma}{2}$, where $\vec{r_B}$ is the Bloch vector and $\sigma$ is the vector of Pauli operators, has the following form,
\begin{equation}\label{channel}
 \varLambda(\rho)=\frac{I_d+(L\vec{r_B}+l)\gamma}{d},
\end{equation}
where ${L}$ is a $(d^{2}-1)\times3$ real matrix and $l$ is a three-dimensional vect

\section{The linear entropy version of the classical correlation}

 Any $d\otimes2$ bipartite  quantum state $\rho_{AB}$ can be written as
(\ref{VB}).
Let $\rho_B=\sum\lambda_i|\phi_i\rangle\langle\phi_i|$ be the spectral
decomposition of $\rho_B$. Then $|V_{B'B}\rangle=\sum\sqrt{\lambda_i}|\phi_i\rangle|\phi_i\rangle$.
One has \cite{prl96220503},
\begin{equation}\nonumber
I^{\leftarrow}_2(\rho_{AB})=\max_{\{p_i,\psi_i\}}\left(S_2[\varLambda(\rho_B)]
-\sum_i p_i S_2[\varLambda(|\psi_i\rangle\langle\psi_i|)]\right),
\end{equation}
where the maximization goes over all possible pure state decompositions of $\rho_B$.
Taking into account (\ref{channel}), we have
\begin{eqnarray*}\nonumber
 S_2[\varLambda(\rho_B)]&=&S_2[\varLambda(\frac{I_2+\vec{r}_B\sigma}{2})]\\
 &=&\frac{2d^2-2d-4(L\vec{r}_B+l)^{T}(L\vec{r}_B+l)}{d^2}.
\end{eqnarray*}

In the Pauli basis, the possible pure state decompositions of $\rho_B$
are represented by all possible sets of probability $\{p_j\}$ and $\vec{r}_j$ such that
$\rho_B=\sum_jp_j\frac{I_2+\vec{r}_j\sigma}{2}$.
Set $\vec{r}_j=\vec{r}_B+\vec{x}_j$. One can easily check that
the calculation of $I^{\leftarrow}_2(\rho_{AB})$ reduces to determine ${p_j,\vec{x}_j}$, subject to the conditions $\sum_jp_j\vec{x}_j=0$ and $|\vec{r}_B+\vec{x}_j|=1$, in the following maximization,
\begin{eqnarray*}\nonumber
 \frac{4}{d^2}\max_{\{p_j,\vec{x}_j\}}\sum_jp_j\vec{x}_j^T L^T L\vec{x}_j.
\end{eqnarray*}
By using the method used in calculating the linear Holevo capacity for qubit channels \cite{prl96220503},
we have the follow Lemma.

{\bf Lemma}~
For arbitrary $d\otimes2$ quantum states,
\begin{eqnarray}\label{lemma}
 I^{\leftarrow}_2(\rho_{AB})=
 \frac{4}{d^2}\lambda_{\max}(L^{T}L)S_2(\rho_{B}),
\end{eqnarray}
where $\lambda_{\max}(L^{T}L)$ stands for the largest eigenvalues of the matrix $L^{T}L$.

By the Lemma, we have corrected a error in \cite{zhihaoma}, where the factor $4/d^2$ in (\ref{lemma}) was missed.

\section{ Analytical formula of quantum discord for rank-2 two-qubit states.}
To get the analytical formula of classical correlation  $I^{\leftarrow}(\rho_{AB})$ under von Neumann entropy from $I^{\leftarrow}_2(\rho_{AB})$
under linear entropy for any bipartite states $\rho_{AB}$, we consider the relations among entanglement of formation, concurrence, tangle, $I^{\leftarrow}(\rho_{AB})$ and $I^{\leftarrow}_2(\rho_{AB})$.
The tangle $\tau(\rho_{AB})$ is defined by
\begin{equation}\nonumber
 \tau(\rho_{AB})=\inf_{\{p_i,|\psi\rangle_i\}}\sum p_iS_2(\rho^i_B),
\end{equation}
where the infimum runs over all pure-state decompositions $\{p_i,|\psi\rangle_i\}$
of $\rho_{AB}$ and $\rho^i_B=Tr_A(|\psi\rangle_i \langle\psi|).$ Due to the convexity, one has
$C^2(\rho_{AB})\leq \tau(\rho_{AB})$ for quantum states.
Generally, $\tau(\rho_{AB})$ is not equal to the square of
the concurrence \cite{pra022309}.

The entanglement of formation $E(|\psi\rangle_{AB})$ \cite{t1,t2,t3} and the concurrence $C(|\psi\rangle_{AB})$ \cite{liming,c2,c3} of a pure state $|\psi\rangle_{AB}$ are defined by
$E(|\psi\rangle_{AB})=S(\rho_A)$ and $C(|\psi\rangle_{AB})=\sqrt{2[1-Tr(\rho^2_A)]}$, respectively.
They are extended to mixed states $\rho_{AB}$ by convex-roof construction,
$E(\rho_{AB})=\inf_{\{p_i,|\psi_i\rangle\}}\sum_ip_iE(|\psi_i\rangle)$,
$C(\rho_{AB})=\inf_{\{p_i,|\psi_i\rangle\}} \sum_i p_i C(|\psi_i\rangle)$,
with the infimum taking over all possible pure state decompositions of $\rho_{AB}$.

For the two-qubit quantum states $\rho_{AB}$,
the entanglement of formation $E_f(\rho_{AB})$ and concurrence $C(\rho_{AB})$
have the following relation \cite{pra90062343}:
\begin{equation}\nonumber\label{FC}
 E_f(\rho_{AB})=h(\frac{1+\sqrt{1-C^2(\rho_{AB})}}{2})
\end{equation}
where $h(x)=-x\log_2(x)-(1-x)\log_2(1-x).$

For a tripartite pure state $|\psi\rangle_{ABC}$, one has the following relations \cite{pra69022309},
\begin{equation}\label{EF}
 E_{f}(\rho_{AC})+I^{\leftarrow}(\rho_{AB})=S(\rho_A).
\end{equation}
In the following we denote $f(x)=h(\frac{1+\sqrt{1-x}}{2})$ for simplicity.

In Ref.\cite{QD1} the authors presented a way to calculate the quantum discord of a rank-2 two-qubit state
$\rho_{AB}=\lambda_0|\phi_0\rangle\langle\phi_0|+\lambda_1|\phi_1\rangle\langle\phi_1|$,
where $|\phi_0\rangle$  and $|\phi_1\rangle$ are the eigenstates of $\rho_{AB}$ with the corresponding
eigenvalues $\lambda_0$ and $\lambda_1$. By attaching a third qubit C the state $\rho_{AB}$ is purified to be
$|\Psi\rangle=\sqrt{\lambda_0}|\phi_0\rangle|0\rangle+\sqrt{\lambda_1}|\phi_1\rangle|1\rangle$.
By local unitary operations one then transforms the
eigenstates $|\phi_0\rangle$ and $|\phi_1\rangle$ simultaneously
to the following forms:
$|\phi_0\rangle=a_0|0\rangle|0\rangle+b_0|\eta\rangle|1\rangle$ and
$|\phi_1\rangle=a_1|1\rangle|0\rangle+b_1|\eta^{\bot}\rangle|1\rangle$,
where $|a_k|^2+|b_k|^2=1$ for $k=0,1$
and $|\eta\rangle=c|0\rangle+d|1\rangle$
is a state which is orthogonal to $|\eta^{\bot}\rangle$ with $|c|^2+|d|^2=1$.
The following results are obtained in \cite{QD1}:
\begin{equation}\label{QD11}\nonumber
C^2(\rho_{BC})=2\lambda_0\lambda_1
\left[|a_0b_1c^{*}-a_1b_0c|^2+2|d|^2(|a_0|^2|b_1|^2+|a_1|^2|b_0|^2)\right]
-2\lambda_0\lambda_1\left|(a_0b_1c^{*}-a_1b_0c)^2-4a_0a_1b_0b_1|d|^2\right|.
\end{equation}
From the relation between $E(\rho_{BC})$ and $C(\rho_{BC})$: $E(\rho_{BC})=h(C^2(\rho_{BC}))$,
one has the entanglement of formation $E(\rho_{BC})$.
From the formula $Q^{\rightarrow}(\rho_{AB})=S(\rho_A)+E(\rho_{BC})-S(\rho_{AB})$,
one obtains the quantum discord $Q^{\rightarrow}(\rho_{AB})$\cite{QD1}.

In the following we present a Theorem which gives an analytical formula of quantum discord for arbitrary rank-2 two-qubit quantum states $\rho_{AB}$.
The formula can be directly calculated for given $\rho_{AB}$ and no purifications are needed.

{\bf Theorem 1}~
For rank-2 two-qubit quantum states $\rho_{AB}$, the quantum discord is given by
\begin{equation}\label{thm}
 Q^{\leftarrow}(\rho_{AB})=S(\rho_B)-S(\rho_{AB})+f(S_2(\rho_A)-I_2^{\leftarrow}(\rho_{AB})).
\end{equation}

{\sf Proof :}~
For two-qubit quantum states $\rho_{AB}$ with rank-2, they have spectral decompositions,
$\rho_{AB}=\lambda_1|\psi\rangle_1\langle\psi|+ \lambda_2|\psi\rangle_2\langle\psi|$,
where $\lambda_i$ and $|\psi\rangle_i$, $i=1,2$,  $\lambda_1+\lambda_2=1$, are respectively the eigenvalues
and eigenvectors. Then the purified tripartite qubit state can be written as
$|\psi\rangle_{ABC}=\sqrt{\lambda_1}|\psi\rangle_1|0\rangle +\sqrt{\lambda_2}|\psi\rangle_2|1\rangle$, satisfying
$\rho_{AB}=Tr_C(|\psi\rangle_{ABC}\langle\psi|)$.
We have the following monogamy relation \cite{prl96220503},
\begin{equation}\label{TAU}
 \tau(\rho_{AC})+I_2^{\leftarrow}(\rho_{AB})=S_2(\rho_A).
\end{equation}

As $\rho_{AC}$ is a two-qubit state, one has $\tau(\rho_{AC})=C^2(\rho_{AC})$ \cite{pra022309}. Moreover,
$S(\rho_A)=E_f(|\psi\rangle_{A|BC})=f(C^2(|\psi\rangle_{A|BC}))=f(S_2(\rho_A))$,
 \begin{eqnarray*}\nonumber
 E_f(\rho_{AC})&=&f(C^2(\rho_{AC}))\\\nonumber
 &=&f\left(S_2(\rho_A)-I^{\leftarrow}_2(\rho_{AB})\right).
 \end{eqnarray*}
 where the fist and second equations are due to (\ref{FC}) and (\ref{TAU}).
From (\ref{EF}), we have
 \begin{eqnarray}\nonumber\label{clc}
 I^{\leftarrow}(\rho_{AB})
 =S(\rho_A)-f\big(S_2(\rho_A)-I^{\leftarrow}_2(\rho_{AB})\big).
 \end{eqnarray}
According to (\ref{discord}), we have the  quantum discord for any rank-2 two-qubit states. $\Box$

Theorem 1 provides an analytical formula (\ref{thm}) of quantum discord in terms of the original  Von Neumann
entropy for arbitrary rank-2 two-qubit quantum states. Besides, the classical correlation $I^{\leftarrow}(\rho_{AB})$ based on the Von Neumann
entropy is also analytically presented.
It should be emphasized that, the analytical formula of quantum discord (\ref{thm})
is only for rank-2 two-qubit quantum states, but the formula for classical correlation (\ref{lemma})
is valid for any $d\otimes 2$ bipartite states with any ranks.
In the following, we give some detailed examples for quantum discords and also classical correlations.

Let us consider the rank-2 of two-qubit Bell-diagonal states,
\begin{equation} \nonumber
\rho=\frac{1}{4}\left(I+\sum_{j=1}^{3}c_i\sigma_j\otimes\sigma_j\right).
\end{equation}
By Theorem 1, we have $S(\rho_A)=1$ and
$S_2(\rho_A)=1$ and $I_2^{\leftarrow}(\rho)=c^2$. Then
$$
I^{\leftarrow}(\rho)=1-f(1-c^2)=\frac{1-c}{2}\log_2(1-c)+\frac{1+c}{2}\log_2(1+c),
$$
which coincides with the result in Ref. \cite{pra77042303}.

{\bf Example 1:} Now consider the following  two-qubit states,
\begin{eqnarray}\nonumber\label{rho1}
\rho_1&=&\frac{2-x}{6}|00\rangle \langle00|+\frac{1+x}{6}|01\rangle \langle01|+\frac{1}{6}|01\rangle \langle10|\\\nonumber
 &+&\frac{1}{6}|10\rangle \langle01|+\frac{1+x}{6}|10\rangle \langle10|+\frac{2-x}{6}|11\rangle \langle11|,\\\nonumber
\end{eqnarray}
where $x\in[0,2]$.
By computation we have $S_2(\rho_B)=1$, and the qubit channel $\varLambda$ is given by
$\varLambda(|0\rangle\langle0|)
=\frac{2-x}{3}|0\rangle\langle0|+\frac{1+x}{3}|1\rangle\langle1|$,
$\varLambda(|0\rangle\langle1|)
=\frac{1}{3}|1\rangle\langle0|$,
$\varLambda(|1\rangle\langle0|)
=\frac{1}{3}|0\rangle\langle1|$
and
$\varLambda(|1\rangle\langle1|)
=\frac{1+x}{3}|0\rangle\langle0|+\frac{2-x}{3}|1\rangle\langle1|.$
Therefore we obtain
\begin{eqnarray}\nonumber
L=\begin{pmatrix}
\frac{1}{3}&0&0&\\
0&-\frac{1}{3}&0&\\
0&0&\frac{1-2x}{3}&\\
  \end{pmatrix}\quad
\end{eqnarray}
and $I_2^{\leftarrow}(\rho_1)=\max_{\{x\in[0.2]\}}\{\frac{1}{9},\frac{(1-2x)^2}{9}\}$,
see Fig.1.

\begin{figure}[htpb]
\centering
\includegraphics[width=7.5cm]{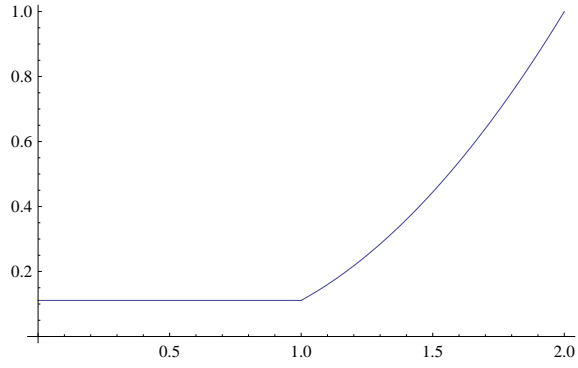}
\caption{{\small The classical correlation $I^{\leftarrow}_2(\rho_1)$ with $x\in[0,2]$ .}}
\label{fig1}
\end{figure}

The rank of $\rho_{1}$ is two when $x=2$. In this case, we have
$S(\rho_{B})=S_2(\rho_{A})=1$ and $S(\rho_{AB})=\log_23-\frac{2}{3}$.
Hence $Q^{\leftarrow}(\rho_{AB})=\frac{5}{3}-\log_23$.

{\bf Example 2:} We calculate now the discord of the Horodecki state \cite{Hstate},
\begin{eqnarray*}\nonumber
  \rho^{H}(p)=p|\varphi^{+}\rangle\langle\varphi^{+}|+(1-p)|00\rangle\langle00|,
\end{eqnarray*}
where $|\varphi^{+}\rangle=\frac{1}{\sqrt{2}}(|01\rangle+|10\rangle)$.
The qubit channel $\varLambda$ can be explicitly calculated: $\varLambda(|0\rangle\langle0|)=\frac{2(1-p)}{2-p}|0\rangle\langle0|
+\frac{p}{2-p}|1\rangle\langle1|,$
$\varLambda(|1\rangle\langle0|)=\sqrt{\frac{p}{2-p}}|1\rangle\langle0|,$
$\varLambda(|0\rangle\langle1|)=\sqrt{\frac{p}{2-p}}|0\rangle\langle1|$
and $\varLambda(|1\rangle\langle1|)=|0\rangle\langle0|.$
By applying Theorem 1, we get the matrix
 \begin{eqnarray} \nonumber
 L=\begin{pmatrix}
 \sqrt{\frac{p}{2-p}}&0&0&\\
 0&-\sqrt{\frac{p}{2-p}}&0&\\
 0&0&-\frac{p}{2-p}&\\
   \end{pmatrix}.\quad
 \end{eqnarray}
It is straightforward to verify that $S_2(\rho^{H}(p)_B)=S_2(\rho^{H}(p)_A)=p(2-p)$
and $S(\rho^{H}(p))=h(p)$.
Thus, the discord of $\rho^{H}(p)$ is given by
\begin{equation} \nonumber
  Q^{\leftarrow}(\rho^{H}(p))
  =h(\frac{p}{2})- h(p)+f(2p(1-p)),
  \end{equation}
see Fig.2.

\begin{figure}[htpb]
\centering
\includegraphics[width=7.5cm]{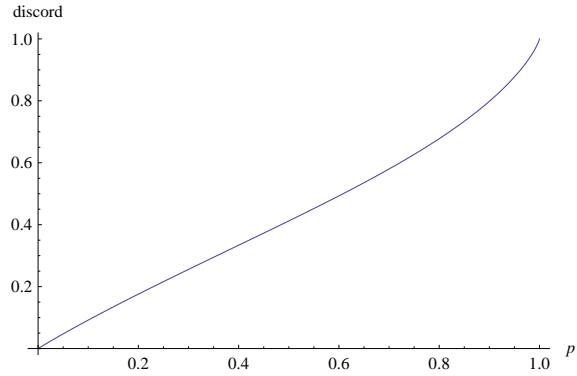}
\caption{{\small The discord of  the Horodecki state $\rho^{H}(p)$.}}
\label{fig2}
\end{figure}

Now we consider some more general rank-2 states,
$$
\rho_2=x|\varphi\rangle\langle\varphi|+(1-x)|\phi\rangle\langle\phi|,
$$
where $|\varphi\rangle=Sin\theta\,|00\rangle+Cos\theta\,|11\rangle,$
$|\phi\rangle=Sin\eta\,|01\rangle+Cos\eta\,|10\rangle,$  $x\in[0,1]$ and $\theta,\eta\in[0,2\pi]$.
Direct computation shows
$L=diag\{L_1,L_2,L_3\}$, where

\begin{eqnarray*}
L_{1}&&=\frac{xSin\theta Cos\theta+(1-x)Sin\eta Cos\eta}{\sqrt{\left[xCos^2\theta+(1-x)Sin^2\eta\right]
[xSin^2\theta+(1-x)Cos^2\eta]}},\\
L_{2}&&=\frac{xSin\theta Cos\theta-(1-x)Sin\eta Cos\eta}{\sqrt{\left[xCos^2\theta+(1-x)Sin^2\eta\right]
[xSin^2\theta+(1-x)Cos^2\eta]}},\\
L_{3}&&=\frac{x^2Sin^2\theta Cos^2\theta-(1-x)^2Sin^2\eta Cos^2\eta}{[xCos^2\theta+(1-x)Sin^2\eta]
[xSin^2\theta+(1-x)Cos^2\eta]},\\
L_4&=&4x(1-x)+x^2Sin^22\theta
+(1-x)^2Sin^22\eta-4x(1-x)Cos^2(\theta-\eta)-2x(1-x)Sin2\theta Sin2\eta,\\
L_5&=&4\left[xSin^2\theta+(1-x)Cos^2\eta\right]
\left[xCos^2\theta+(1-x)Sin^2\eta\right],
\end{eqnarray*}

$S(\rho_B)=h\left(xSin^2\theta+(1-x)Cos^2\eta\right),$
$S(\rho_2)=h(x),$
and
$S_2(\rho_{A})=L_4.$
Therefore we obtain

\begin{equation} \nonumber
  Q^{\leftarrow}(\rho_{AB})
  =h\left(xSin^2\theta+(1-x)Cos^2\eta\right)-h(x)
  +f(L_4-\max_{\{i=1,2,3\}}\{L^2_i\}L_5).
  \end{equation}

The Horodecki state $\rho^{H}(p)$ is a special case of $\rho_2$ at $\theta=\frac{\pi}{2}$,
$\eta=\frac{\pi}{4}$ and $x=1-p$.

\section{Conclusion}

By analyzing the relations among the entanglement of formation, concurrence, tangle, linear entropy based classical correlation and
von Neumann entropy based classical correlation, we have derived the analytical formulae of classical correlations under  linear entropic  for arbitrary $d\otimes 2$ states and under von Neumann entropic  for arbitrary $2\otimes 2$ rank-2 states.
From the von Neumann entropy based classical correlation, we have presented explicit formula of quantum discord for arbitrary rank-2 two-qubit quantum states.
If one can further get the relation between $\tau(\rho_{AB})$ and
$E(\rho_{AB})$ for rank-2 $d\otimes 2$ systems, it would be possible to compute the quantum discord for rank-2 $d\otimes 2$ states. And if one is able to get the relation between $\tau(\rho_{AB})$  and
$E(\rho_{AB})$ for $4\otimes 2$ systems, maybe one can compute the discord for any two-qubit states.
However, for the rank-2 mixed states $\rho_{AB}$, the corresponding entanglement of formation satisfies the inequality
$E(\rho_{AB})\leq f(\tau)$ \cite{pra022309}. The tangle $\tau(\rho_{AB})$ is not, in general, equal to the square of concurrence $C^2(\rho_{AB})$.
It is of difficulty to calculate the discord of any rank-2 $d\otimes2$ quantum states and any two-qubit states.

\bigskip
\noindent{\bf Acknowledgments}\, \,
We thank Ming Li, Huihui Qin and Tinggui Zhang for helpful discussions.
This work is supported by NSFC under numbers 11675113 and 11605083, and NSF of Beijing under No. KZ201810028042.

\end{document}